\documentclass[aps,prb,reprint,groupedaddress,showpacs]{revtex4-1}

\usepackage{graphicx}
\usepackage{amsmath}
\usepackage{amsfonts}
\usepackage{color}

\begin{document}


\title{Theoretical Studies of Graphene Nanoribbon Quantum Dot Qubits}


\author{Chih-Chieh Chen}
\affiliation{Department of Physics, University of Illinois at Urbana-Champaign, Urbana, Illinois 61801, USA}
\affiliation{Research Center for Applied Sciences, Academic Sinica, Taipei, 11529 Taiwan}

\author{Yia-Chung Chang}
\email[]{yiachang@gate.sinica.edu.tw}
\affiliation{Research Center for Applied Sciences, Academic Sinica, Taipei, 11529 Taiwan}
\affiliation{Department of Physics, National Cheng-Kung University, Tainan, 70101 Taiwan}


\date{\today}

\begin{abstract}
Graphene nanoribbon quantum dot qubits have been proposed as promising candidates for
quantum computing applications to overcome the spin-decoherence problems associated with
typical semiconductor (e.g. GaAs) quantum dot qubits. We perform theoretical studies of the electronic structures of graphene
nanoribbon quantum dots by solving the Dirac equation with appropriate boundary conditions. We then evaluate
the exchange splitting based on an unrestricted Hartree-Fock method for the Dirac particles. The electronic wave function and long-range exchange coupling due to the Klein tunneling and the Coulomb interaction are calculated for various gate
configurations. It is found that the exchange coupling between qubits can be significantly enhanced by the Klein tunneling effect.
The implications of our results for practical qubit construction and operation are discussed.
\end{abstract}


\maketitle





\section{Introduction}

Despite rapid progress in computer technology, there are still computational problems that are
difficult to solve for any known algorithm that uses modern computers. However, the theory that
describes physics on atomic length-scales, quantum mechanics, suggests a new way to attack
hard computational problems in a more efficient way. A computer using operations involving
the entanglement of quantum states is known as a quantum computer. \cite{Deutsch1985} A number of algorithms proposed for
quantum computers are expected to solve many classically hard problems. \cite{nielsen2010quantum} For example,
Shor's algorithm \cite{Shor1997} can efficiently solve the prime factorization problem, which is difficult to
solve even with state-of-the-art classical algorithms and computers.
The quantum analog of the bit -- the fundamental information storage unit in classical
computers -- is called the qubit. Many physical realizations of  qubits for quantum computers
are being developed, including semiconductor, superconductor, nuclear, and optical qubit
systems. The mature semiconductor manufacturing industry today offers several advantages in
the construction of semiconductor qubits. In particular, the nanoscale semiconductor structure
known as the quantum dot has been proposed as a possible realization of the semiconductor
qubit. \cite{LossDiVincenzo1998}

Semiconductor qubits constructed from typical semiconductor (e.g. GaAs) quantum dots have limited usefulness, because
quantum information stored in the device can be lost due to spin-decoherence. Sources of
spin-decoherence in semiconductors include the spin-orbit, hyperfine, and electron-phonon interactions.
{Graphene, \cite{RevModPhys.81.109,Kotov2012,DasSarma2011} a two-dimensional lattice structure formed by carbon, is a promising material for
avoiding spin-decoherence. \cite{trauzettel2007spin} Graphene has not only weak spin-orbit coupling, but also a
negligible hyperfine interaction, since carbon-12 has zero nuclear spin. These advantages of
graphene have driven researchers towards developing a graphene-based qubit. One proposed
model for the graphene qubit is the graphene nanoribbon (GNB) quantum dot, \cite{trauzettel2007spin,BreyFertig2006,Gucclu2014} which has been
experimentally realized by several groups. \cite{Han2007,Li2008,Stampfer2009,Guttinger2012,Liu2010,Wei2013}}

Previous theoretical efforts to model GNB quantum dots \cite{trauzettel2007spin} offer only
estimates of the electron-electron interaction and many-particle effects in graphene.
Consequently, it is not clear if the predicted long-range Heisenberg exchange coupling between
the dots---which is necessary for qubit operations in universal quantum
computing---can be achieved in practice. Nor are the effects of gate voltage changes on the
electronic structure and the exchange coupling in the multi-electron regime well understood.
Realistic models of graphene qubit operation require a better understanding of these effects. \cite{LossDiVincenzo1998}
To understand how the GNB quantum dot qubit functions in real
applications, we study a more complete model of the graphene qubit using a numerical
approach. {  To study the exchange coupling between two graphene qubits, we use an unrestricted Hartree-Fock method with a generalized-valence-bond (GVB) wave function. \cite{Ostlund1996,Hu2000,Yannouleas2001,Fang2002} This is a double-Slater-determinant approach, which takes into account the correlation effect due to charge separation and is suitable for describing diatomic molecules. In contrast, the conventional Hartree-Fock method, which uses factorized single-particle product states (i.e. a single determinant) cannot describe the correlation effect. The GVB method allows us to capture the physics of entangled states (crucial for quantum computing), which cannot be described by the single-determinant Hartree-Fock method. \cite{Horodecki2009} The computational cost of the GVB method is much lower than that of the configuration interaction calculations. \cite{Ostlund1996,Dutta2008} Thus, it is easier to analyze the exchange coupling between two qubits in various gate configurations with the GVB method.} In this work, we employ this numerical scheme and the Dirac equation to provide a realistic simulation of a GNB quantum dot qubit.

\section{Method}

\subsection{One-particle problem}

Consider a graphene nanoribbon with width $W$ and length $L$ and armchair boundary conditions. Let the x-axis be the direction across the width of the nanoribbon and the y-axis be the direction parallel to the length of the ribbon. { The behavior of an electron with energy close to the Fermi level in this system can be described by the Dirac equation \cite{Semenoff1984,DiVincenzo1984,trauzettel2007spin,BreyFertig2006,RevModPhys.81.109}

\begin{eqnarray}
&& H_1 |\psi \rangle = \epsilon   | \psi \rangle \\
&& H_1 =  -i \hbar v
\begin{pmatrix}
\sigma_x  \partial_x + \sigma_y  \partial_y & 0 \\
0 & -\sigma_x  \partial_x + \sigma_y  \partial_y  \\
\end{pmatrix}
 +V(y) ,\notag
\end{eqnarray}
where $\hbar$ is Planck's constant, $v$ is the Fermi velocity of graphene, $\sigma_x,\sigma_y$ are Pauli matrices for the pseudospin describing two sublattices of graphene, $\partial_x,\partial_y$ are partial derivatives, and $V(y)$ is the electrical confining potential along the y-axis.} $|\psi \rangle$ is a 4-component spinor in the form
\begin{eqnarray}
|\psi \rangle =
\begin{pmatrix}
\psi (K,A)\\
\psi (K,B)\\
-\psi(K',A)\\
-\psi(K',B)\\
\end{pmatrix},
\end{eqnarray}
where $K,K'$ label the two valleys in the Brillouin zone of graphene, $A,B$ label the two sublattices of graphene, and $\psi$ is the envelope function.

We expand the electron envelope function within a basis set as follows
\begin{eqnarray}
 |\psi \rangle &=& \sum_{m,s,n} \phi^{m,n}_s   |\psi^{m,n}_s \rangle, \\
 \langle x,y  |\psi^{m,n}_s \rangle &=&
\frac{1}{\sqrt{2WL}}  \begin{pmatrix}
\chi_s e^{iq_nx} \\
\chi_s e^{-iq_nx}  \\
\end{pmatrix}
 f_m (y) ,
\end{eqnarray}
 where $\{ f_m(y) \}$ is a set of basis functions of variable $y$, $s=A,B$ labels the two components of the pseudospin describing two sublattices. The basis vectors for the two-component pseudospinors are
\begin{eqnarray}
&\chi_A &=
\begin{pmatrix}
1 \\
0 \\
\end{pmatrix} \\
&\chi_B &=
\begin{pmatrix}
0 \\
1 \\
\end{pmatrix}.
\end{eqnarray}
The boundary conditions are
\begin{eqnarray}
&\langle x,y|\psi \rangle |_{x=0} &=
\begin{pmatrix}
 0 & {\bf I}_{2} \\
 {\bf I}_{2} & 0 \\
\end{pmatrix}
\langle x,y |\psi \rangle |_{x=0} \label{bc1} \\
&\langle x,y|\psi \rangle |_{x=W} &=
\begin{pmatrix}
 0 & e^{+ i\frac{2\pi\mu}{3} }{\bf I}_{2} \\
 e^{ -i\frac{2\pi\mu}{3} }{\bf I}_{2} & 0 \\
\end{pmatrix}
\langle x,y |\psi  \rangle |_{x=W} , \label{bc2}
\end{eqnarray}
where $\mu =\pm 1, 0$ is a constant determined by the termination of the ribbon edge. $\mu =\pm 1$ defines the semiconducting boundary condition. The symbol ${\bf I}_{2}$ in Eqs. (\ref{bc1}) and (\ref{bc2}) denotes a $2\times2$ identity matrix in pseudospin space. {Imposing the  semiconducting boundary conditions on the  basis functions leads to quantization of the electronic states in the x-direction
\begin{eqnarray}
 &q_n &=\frac{\pi}{W}(n + \frac{\mu}{3}) , n \in \mathbb{Z} \\
 &&= (3n + \mu )q_0 ,
\end{eqnarray}
where the characteristic momentum scale $q_0 =  \frac{\pi}{3W}$ is defined by the width of the ribbon $W$. In this work, $1/q_0$ is the characteristic length scale and $\hbar v q_0$ is a characteristic energy scale. Throughout this paper, we consider only the condition with $n=0$ and $\mu=+1$, since the ribbon is narrow enough such that the energies of higher confined states are outside the range of interest.}

The basis functions $f_m(y)$ are chosen to be sinusoidal functions confined within the interval $[0,L]$,
\begin{eqnarray}
f_m(y) &=& \sqrt{2}\sin (\frac{\pi m y}{L}) .
\end{eqnarray}
The matrix elements of the one-particle Hamiltonian and overlap can then be written down analytically. {The Dirac equation is cast into the form of a generalized eigenvalue problem
\begin{eqnarray}
\sum_{ms} \langle \psi^{m',n}_{s'} | H_1 | \psi^{m,n}_{s} \rangle  \phi^{m,n}_s = \epsilon \sum_{ms} \langle \psi^{m',n}_{s'}  | \psi^{m,n}_s \rangle \phi^{m,n}_s , \notag \\
\end{eqnarray}
which is solved numerically.}

\subsection{Two-particle problem}

To evaluate the exchange coupling between two electrons separately located in two neighboring GNB quantum dots, we need to consider the mutual Coulomb interaction and exchange term between them.  The Coulomb interaction in two-dimension is given by {
\begin{equation}
 v_{ee} = (\frac{e^2}{4\pi\epsilon \hbar v }) \hbar v
  \frac{1}{\sqrt{(x_1-x_2)^2+(y_1-y_2)^2}}, \notag
\end{equation}}
where $(\frac{e^2 }{4\pi\epsilon \hbar v })$ is a dimensionless Coulomb parameter, which can be viewed as the fine-structure constant of graphene. {One expects $\frac{e^2 }{4\pi\epsilon \hbar v }=2.2$ or smaller for a suspended graphene.\cite{Hwang2012,Reed2010,Kotov2012} In this work we use $\frac{e^2 }{4\pi\epsilon \hbar v }=1.43$ for graphene on quartz substrate. \cite{Hwang2012}}

We adopt the unrestricted Hartree-Fock method with generalized-valence-bond wave function \cite{Fang2002} to solve the two-particle problem. Because the spin-orbit interaction is omitted here, the total wavefunction can be written as the product of the spatial wavefunction, denoted by $\Psi_+ (\Psi_-)$, and the corresponding two-particle spinor for the singlet (triplet) state. The spatial wave function $\Psi_+ (\Psi_-)$ of the spin singlet (triplet) must be symmetric (antisymmetric) with respect to the exchange of two particles to make the total wavefunction antisymmetric. The spatial wavefunctions take the form
\begin{eqnarray}
 | \Psi_\pm \rangle =
\frac{1}{\sqrt{2(1\pm S^2)}}( | \psi_L , \psi_R \rangle \pm | \psi_R , \psi_L \rangle),
\end{eqnarray}
where $\psi_L$ ($\psi_R$) denotes a four-component single-particle wavefunction (for the Dirac particle with two bands and two valleys at K and K') localized at the left (right) quantum dot, and $S=\langle \psi_L | \phi_R \rangle$, which is enhanced when the Klein tunneling condition is met.  The two-particle Hamiltonian can be written as
\begin{eqnarray}
H = H_1 \otimes 1 + 1\otimes H_1 + v_{ee},
\end{eqnarray}
where $H_1$ is the single-particle Hamiltonian for a Dirac particle as defined in Eq.~(1).
The two-particle Schrodinger equation is
\begin{eqnarray}
& H | \Psi_\pm \rangle  &= E | \Psi_\pm \rangle .
\end{eqnarray}
$\psi_L$ ($\psi_R$) appearing in $| \Psi_\pm \rangle$ is expanded in a non-orthonormal basis set $\{|\nu \rangle\}$ as defined in Eq.~(4) with $\nu$ being a composite index for $(m,n,s)$.
In each iteration with a given $\psi_R$, we write $| \psi_L \rangle=\sum_{\nu}C^L_{\nu}|\nu \rangle$ and solve for the expansion coefficients of $C^L_{\nu}$ according to the following projected Scr\"{o}dinger equation for $| \Psi_\pm \rangle$.
\begin{eqnarray}
& \langle \nu',\psi_R|H| \Psi_\pm \rangle  &= E \langle \nu',\psi_R | \Psi_\pm \rangle ,
\end{eqnarray}
where $\langle\nu' , \psi_R|$ denotes a product state of basis state $\langle \nu'|$ and $\langle \psi_R|$.

The generalized eigenvalue problem to be solved becomes
\begin{eqnarray}
\sum_{\nu} \langle \nu' | H_{GVB} |\nu \rangle C^L_{\nu} = E \sum_{\nu} \langle \nu' | S_{GVB} |\nu \rangle C^L_{\nu},
\end{eqnarray}
where the Hamiltonian and overlap matrices elements are
\begin{eqnarray}
\langle \nu' | H_{GVB} |\nu \rangle &=&  \langle \nu' |H_1| \nu \rangle  + \langle \nu' | \nu \rangle   \langle \psi_R |H_1| \psi_R \rangle \\
& \pm & \langle \nu' |H_1| \psi_R \rangle \langle \psi_R |\nu \rangle \pm \langle \nu' | \psi_R \rangle \langle \psi_R |H_1| \nu \rangle   \notag \\
&+& \langle \nu',\psi_R | v_{ee} | \nu, \psi_R \rangle  \pm  \langle \nu',\psi_R | v_{ee} |\psi_R , \nu \rangle \notag \\
 \langle \nu' | S_{GVB} |\nu \rangle &=& \langle \nu' | \nu\rangle   \pm \langle \nu' | \psi_R \rangle  \langle \psi_R |\nu\rangle .
\end{eqnarray}
The iteration continues until self-consistency is reached. For the triplet state, we carry out the reorthonormalization and projection procedure described in \textcite{Fang2002} to resolve the linear-dependence problem in the generalized eigenvalue problem.

\subsection{The double well model}

\begin{figure}
\includegraphics[trim = 0mm 0mm 0mm 0mm,width=0.4\textwidth]{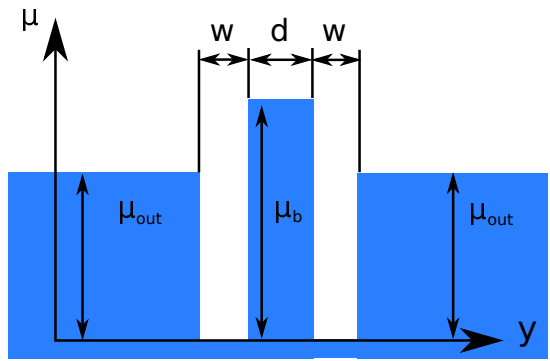}
\caption{Double well potential profile along the graphene ribbon length. The width of each well is $w$, the well-to-well separation is $d$,  the { barrier  height outside the double wells is $\mu_{out}$, the height of the middle potential barrier is $\mu_b$, and the potential at the bottom of the well is set to zero.}\label{fig1}}
\end{figure}

We model a GNB double-dot system by a double square well potential in the y-direction, as shown in Fig.~\ref{fig1}. Unless specified, we use the following parameters throughout this work. The physical parameters used in the model are those reported in the experimental study by \textcite{Liu2010} The width of the ribbon is {$W=20 (nm)\approx 1q_0^{-1}$}, and hence our characteristic energy is {$\hbar v q_0\approx 32.9 (meV)$}. The length of the ribbon is {$L=800 (nm)\approx 40q_0^{-1}$}. The width of each dot (i.e., the width of each confining well) is $w$. The separation between the dots (i.e., the width of the potential barrier between two dots) is $d$. {The potential heights of the barrier and outside region are given by $\mu_b$ and $\mu_{out}$, respectively. The potential at the bottom of the well is set to zero.} We use $\mu_{out}=1.5 \hbar vq_0 =49.4 (meV)$ as suggested by previous theoretical work. \cite{trauzettel2007spin} The Fermi energy is fixed at $E_F=1 \hbar vq_0$. The number of sinusoidal basis functions used is 50.

\section{Results and Discussion}

\subsection{{Single-particle solutions}}

\begin{figure}
\includegraphics[trim = 20mm 0mm 80mm 0mm,width=0.45\textwidth]{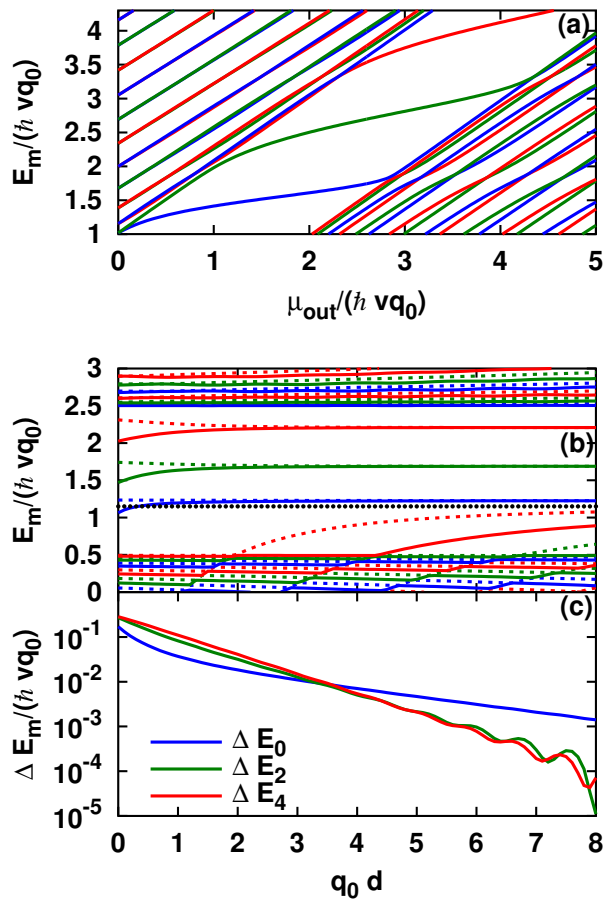}
\caption{ (a) Single particle energy levels as functions of confining potential height for a single well. The well width is $w=2q_0^{-1}$. The conducting states (states forming the left upper triangle) and the Klein tunneling states (states forming the right lower triangle) are also shown in our calculation. (b) Single particle energy levels as functions of inter-well distance for a double well. The width of the wells is $w=4q_0^{-1}$. The ground state (solid blue) is generally above but very close to the top of the barrier valence band (dotted black), and hence is a {Klein} tunneling assisted state. (c) Energy splittings ($\Delta E_m= E_{m+1}-E_m$) as functions of inter-well distance for a double well. \label{fig2}}
\end{figure}

We first examine the single-particle behavior of single and double potential wells. Figure \ref{fig2}(a) shows the single-particle energy levels as functions of confining barrier height $\mu_{out}$ for a single potential well, where the width of the well is $w=2q_0^{-1}$ and the length of the ribbon is $L=16 q_0^{-1}$. { The left upper triangle region contains many slanted lines which describe the discretized states of the conduction bands derived from the GNB regions outside the well. Similarly, the right lower triangle region also contains many slanted lines which describe the discretized states of the GNB valence bands. The discretization is a result of quantum confinement due to the finite length ($L$) of GNB considered. The two regions are separated by a constant $2 \hbar v q_0$, which corresponds to the band gap of GNB. In the gap, there are 3 quantized levels, which are the quantum confined states derived from the middle well. The dependence of these energy levels on the barrier height $\mu_{out}$ matches the results obtained by solving the transcendental equation as described in \textcite{trauzettel2007spin}. The good agreement validates our numerical procedure. It should be noted that our numerical method based on the expansion within a nearly complete basis set can handle not only the simple case of a GNB quantum dot defined by a square well but also arbitrary potential profile along the $y$ axis. This makes it easy to extend to double wells and the two-particle problem for finding the exchange coupling between two neighboring qubits.}

Figure~\ref{fig2}(b) shows energy levels as functions of inter-well separation, $d$ for a double square well. The width of the wells is $w=4q_0^{-1}$. This figure involves some high-energy excited states, so the number of sinusoidal basis functions used in this calculation is enlarged to 100 for higher accuracy. These energy levels come in pairs (indicated by the same color but different curve types), corresponding to two split levels associated with the inter-well coupling of one energy level in the left dot and the corresponding one in the right. The amount of energy splitting reflects the coupling strength of the two states located in the two dots, which would be degenerate in energy in the absence of the inter-well coupling. {Above the energy of the valence band maximum (VBM) of the barrier (black dotted line), there are three pairs of bound states indicated by blue, green, and red colors.  The ground state (blue) pair  is very close to the VBM of the barrier, and one expects to see enhanced inter-well coupling due to the Klein-tunneling effect.}

Figure~\ref{fig2}(c) shows the energy splittings due to inter-well coupling ($\Delta E_m=E_{m+1}-E_m$) as functions of the well-to-well separation, $d$ for the same double square well potential in semi-log scale. $E_0$ is the ground state energy, $E_1$ is the 1st-excited state energy ...etc. The curves are almost linear in semi-log scale, indicating that the energy splittings decay exponentially with the well-to-well separation. For small separation the ground state splitting, $\Delta E_0$ is smaller than the excited state splittings $\Delta E_m, m=2,4$. However, for large separation ($q_0d>4$) the ground state splitting becomes larger than the excited state splittings, indicating the ground state splitting has a smaller decay rate comparing to the excited state splittings. This suggests that the inter-well coupling is enhanced by the Klein tunneling for the ground state pair, in qualitative agreement with the prediction in Ref.~\onlinecite{trauzettel2007spin} based on simple estimation.

\subsection{{Effects of barrier height on two-particle solutions}}

\begin{figure}
\includegraphics[trim = 35mm 0mm 45mm 0mm,width=0.5\textwidth]{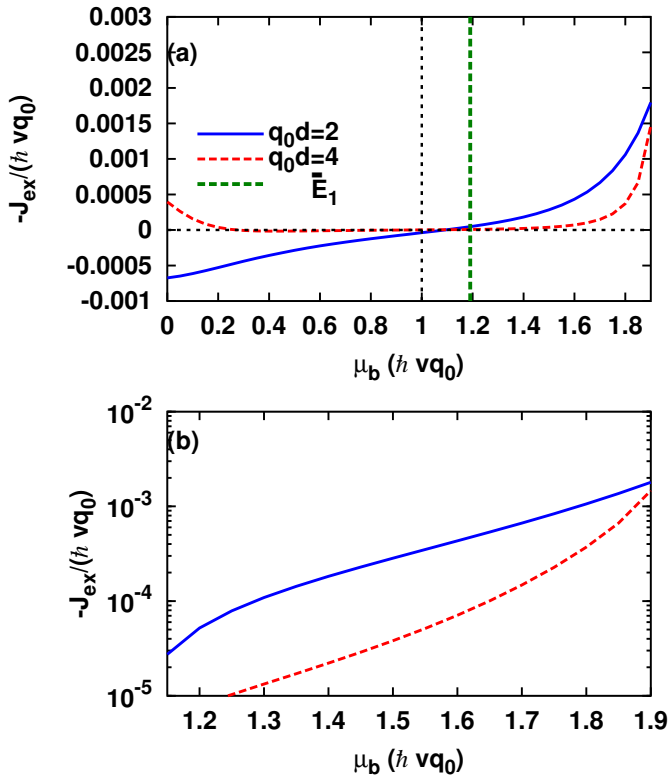}
\caption{Negative exchange coupling $-J_{ex}$ as a function of the inter-well barrier potential height $\mu_b$ for $q_0d=2$ (blue solid) and $q_0d=4$ (red dashed). The width of the wells is $w=4q_0^{-1}$. (a) Linear plot. (b) Semi-log plot. \label{fig3}}
\end{figure}

{ Here, we study the effects of barrier height on the exchange coupling ($J_{ex}=E_{triplet}-E_{singlet}$) between two electrons in the GNB double well. Figure~\ref{fig3}(a) shows  $-J_{ex}$ as a function of barrier potential height $\mu_b$. For small barrier heights, $-J_{ex}$ can be either negative (for $q_0d=2$, blue solid) or positive (for $q_0d=4$, red dashed) depending on the well-to-well separation.} For small barrier heights, the potential profile behaves like a single confining well instead of a double well. A singlet-triplet ground state transition is expected for barrier heights lower than a critical value, which shall be discussed further in Section \ref{distance}. In this section we focus on the regime in which the barrier height is larger than or equal to the critical value. In our model, the critical value can be estimated by comparing the barrier height with both the single-particle ground state energy and the bottom of the conduction bands associated with the wells. In Fig.~\ref{fig3}(a), the green dashed line marks the point when $\mu_b$ crosses $\langle H_1 \rangle$, the expectation value of $H_1$ in the triplet solution for $q_0d=2$, which has a weak dependence on $\mu_b$. We label this point by $\bar E_1$. $\bar E_1$ happens to be almost the same as the average of $\langle H_1 \rangle$ over $\mu_b$ for $\mu_b$ from 0 to 2$\hbar vq_0$.
The black dashed line indicates where the barrier height equals to the bottom of the conduction bands of the wells, which is at $1\hbar vq_0$. {The critical value $\approx 1.1\hbar vq_0$, which lies between $1\hbar vq_0$ and $\bar E_1$.}
For $\mu_b$ larger than the critical value {($\approx 1.1\hbar vq_0$) $-J_{ex}$} is always positive, and it increases monotonically up to $\mu_b= 1.9\hbar vq_0$.

Figure~\ref{fig3}(b) shows $-J_{ex}$ for $\mu_b > 1.1\hbar vq_0$ on a semi-log scale. The magnitude of  $-J_{ex}$ grows exponentially for $q_0d=2$, and super-exponentially for $q_0d=4$. The coupling for $q_0d=4$ can be almost as large as the case for $q_0d=2$ as $\mu_b$ reaches $ 1.9\hbar vq_0 $. The exchange coupling for $q_0d=2$ (blue solid) is linear in the semi-log plot, which indicates an exponential growth. For a longer well-to-well separation with $q_0d=4$, one can see super-exponential growth of the exchange coupling. For $q_0d=4$ and $\mu_b >1.92 \hbar vq_0$, the electrons in the singlet state are in the first-excited single-particle state of both dots, while the electrons in the triplet state are still in the single-particle ground state. The exchange coupling can not be defined in this case.

 \begin{table}
 \caption{Singlet total energy $E_{singlet}$, triplet total energy $E_{triplet}$, and triplet single particle energy $E_1$ for some selected inter-dot distance $d$ and barrier height $\mu_b$.\label{table1}}
 \begin{ruledtabular}
 \begin{tabular}{c c c c c}
 $q_0d$ & $\mu_b/\hbar vq_0$ & $E_{singlet}/\hbar vq_0$ & $E_{triplet}/\hbar vq_0$ & $E_1/\hbar vq_0$\\ \hline
 2 & 1.5 & 2.63897  & 2.63869 & 1.20982 \\
 2 & 1.9 & 2.65891 &   2.65711 & 1.21937 \\
 4 & 1.5 & 2.57493 &  2.57489  & 1.20796 \\
 4 & 1.9 &  2.59815 &   2.59668 &   1.21833   \\
 8 & 1.5 & 2.51309 &  2.51309  & 1.20725 \\
 8 & 1.9 &  2.53504  &  2.53501 & 1.21845 \\
 \end{tabular}
 \end{ruledtabular}
 \end{table}

The super-exponential growth of $-J_{ex}$ as $\mu_b$ increases for larger well-to-well separation as shown in Fig.~\ref{fig3}(b) is a special characteristic of the GNB quantum dot qubit. The overlap between the wave functions of the electrons in the left dot and the right dot is expected to be enhanced by the Klein tunnelling of Dirac particles when the valence band maximum of the barrier is close to the energies of conduction band states in the wells. \cite{trauzettel2007spin} This long-distance coupling of Dirac particles is suggested as a possible advantage of the GNB quantum dot qubit over qubits in conventional systems. For the GNB qubit, the exchange coupling for the long distance case ($q_0d=4$) is almost as large as that in the short distance case ($q_0d=2$) as the barrier height approaches $\mu_b=2\hbar vq_0$. This implies that the valence band maximum in the barrier region is approaching the bottom of conduction band in the well. This result lends support to the proposal in \textcite{trauzettel2007spin}.

\begin{figure}
\includegraphics[trim = 35mm 0mm 35mm 0mm,width=0.33\textwidth]{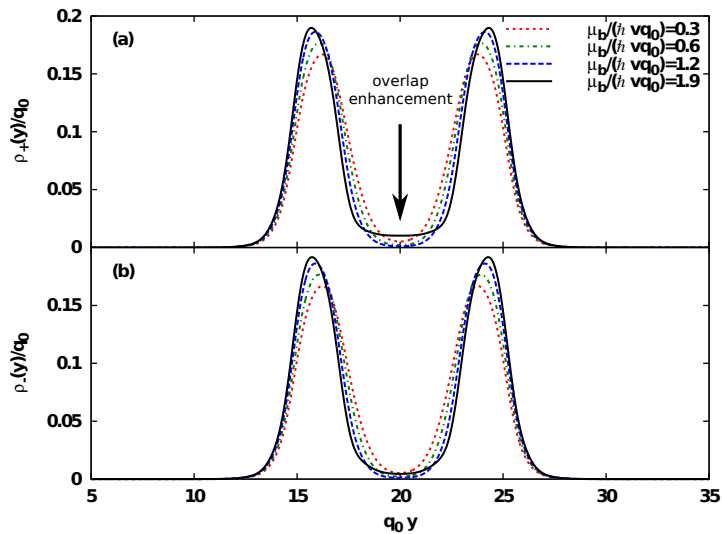}
\caption{Projected charge densities $\rho_\pm(y)$ along the ribbon for the case $q_0d=4$ for (a) singlet state and (b) triplet state for various barrier potential strengths. $\mu_b=0.3\hbar vq_0$ (red dotted), $\mu_b=0.6\hbar vq_0$ (green dash-dot), $\mu_b=1.2\hbar vq_0$ (blue dashed), $\mu_b=1.9\hbar vq_0$ (black solid). The density in the barrier region decreases as $\mu_b$ increasing from zero to $0.6$ and $1.2 \hbar vq_0$, but increases as $\mu_b$ further increases to a higher value $1.9\hbar vq_0$. The increment of density in the barrier region is significantly larger for the singlet state than that of the triplet state. The width of the wells is $w=4q_0^{-1}$.\label{fig4}}
\end{figure}

{This overlap enhancement can be illustrated by examining the projected charge density, defined as
\begin{equation} \rho_\pm (y_1)=\int {\Psi}_\pm(x_1,y_1;x_2,y_2)\cdot \Psi_\pm(x_1,y_1;x_2,y_2) dx_1dx_2dy_2, \end{equation}
 where $\Psi_\pm$ is a four-dimensional vector defined in Eq.~(13). { The projected charge density can be written explicitly as
\begin{eqnarray} \rho_\pm (y_1)&=&\int dx_1 [|\psi_L(x_1,y_1)|^2/2+|\psi_R(x_1,y_1)|^2/2 \nonumber \\
&\pm & S \psi_L(x_1,y_1)\psi_R(x_1,y_1)]/(1\pm S^2). \end{eqnarray} 
For all cases, the $\rho_\pm (y)$ is dominated by the sum of the first two terms, since the overlap $S$ is small. We plot $\rho_\pm (y)$ as a function of $y$ in Fig.~\ref{fig4} for singlet and triplet states for various barrier heights. The electron density is mainly localized in the two potential wells. The charge density between two wells decreases as the barrier potential is raised from $\mu_b=0.3\hbar vq_0$ (red dotted) to $0.6\hbar vq_0$ (green dash-dot) and $1.2\hbar vq_0$ (blue dashed). However, as the barrier height is raised to a higher value $\mu_b=1.9\hbar vq_0$ (black solid), the charge density between the wells becomes significantly larger for the singlet state as seen in Fig.~4(a). In contrast, there is much smaller change in charge density for the triplet state for the corresponding change in barrier height. The enhanced charge density at the middle for the singlet is caused by the long tails of $|\psi_L(x_1,y_1)|^2$ and $|\psi_R(x_1,y_1)|^2$ extending into the opposite well, which result from the Klein tunneling effect. The long tail is enhanced (reduced) for the singlet (triplet) state, which is driven by the positive (negative) exchange term.}

This counter-intuitive behavior due to Klein tunneling is one of the special characteristics of Dirac particles. As the barrier height approaches $\mu_b=2\hbar vq_0$, the VBM of the barrier is aligned with the conduction band minimum in the well. This leads to an enhancement of the overlap between the two electrons. This result is consistent with the result shown in Fig.~\ref{fig3}.

\subsection{{Effects of inter-dot distance on two-particle solutions}\label{distance}}

\begin{figure}
\includegraphics[trim = 30mm 0mm 40mm 0mm,width=0.45\textwidth]{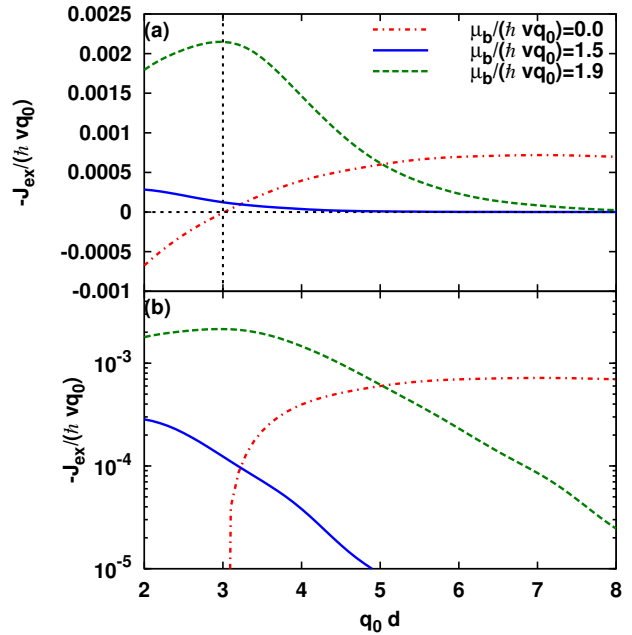}
\caption{Negative exchange coupling ($-J_{ex}$) as a function of well-to-well distance $d$ for various barrier heights in (a) linear scale and (b) semi-log scale. The width of the wells is $w=4q_0^{-1}$. $\mu_b=0$ (red dash-dot), $\mu_b=1.5\hbar vq_0$ (blue solid), $\mu_b=1.9\hbar vq_0$ (green dashed). For zero barrier, the singlet-triplet ground state transition occurs roughly at critical distance {with $q_0d_c\approx 3$} (black vertical dashed). For finite barrier,  $-J_{ex}$ decays exponentially for $q_0d>3$.\label{fig5}}
\end{figure}

Figure~\ref{fig5}(a) shows  $-J_{ex}$  as a function of inter-dot distance for different barrier heights. In the absence of a barrier (red dash-dot),  $-J_{ex}$  starts with negative values and increases to positive values for $q_0d>3$. For $\mu_b=1.5\hbar vq_0$ and $\mu_b=1.9\hbar vq_0$,  $-J_{ex}$  starts with a positive value, and decays exponentially for $q_0d>3$, as shown in Fig.~\ref{fig5}(b).

For $\mu_b=0$, there is no barrier and we only have one confining potential well. For $\mu_b=0$, increasing $d$ is the same as increasing the width of a single potential well (which equals to $d+2w$, as shown in Fig.~1). {This situation has been studied by using various first-principle calculations, as summarized in Ref.~\onlinecite{Rayne2011}. The ground state is a singlet for a short ribbon, and a triplet for a long ribbon. Our result is consistent with the previous studies for this limiting case. In particular, the red dash-dot line is similar to Fig. 2(a) in Ref.~\onlinecite{Rayne2011}. The change of sign of $J_{ex}$ as $q_0d$ varies has a simple physical explanation. Whether the ground state is the singlet or triplet state depends the relative strength of the kinetic energy and the potential energy due to mutual Coulomb interaction. Similar to interacting electrons in a jellium model, the kinetic energy of the system scales like $1/r_s^2$ and potential energy scales like $1/r_s$, where $r_s$ denotes the average distance between electrons in the system. For small inter-dot separation, the kinetic energy dominates and the singlet state has lower total energy since its wavefunction has less spatial variation (due to symmetric behavior) as compared to the the triplet state. A the separation gets larger, the potential energy dominates and the total energy of the singlet state becomes higher, since it has more charge piling toward the center as compared to the the triplet state as illustrated in Fig.~4.} In our calculation, the singlet-triplet ground state transition occurs roughly at the critical distance {$d_c\approx 3q_0^{-1}$}, which is labeled by the black vertical dashed line. For larger barrier heights, the singlet state has a larger density at the central barrier region and hence has higher Coulomb energy. This is why the triplet state is the ground state and  $-J_{ex}$ is always positive for $\mu_b=1.5\hbar vq_0$ and $\mu_b=1.9\hbar vq_0$.

For medium barrier height $\mu_b=1.5\hbar vq_0$, the exchange coupling decreases exponentially when the inter-dot distance increases, as shown in Fig.~\ref{fig5}. For higher barrier height $\mu_b=1.9\hbar vq_0$, where Klein paradox assisted tunneling occurs,  $-J_{ex}$  is generally larger than that for $\mu_b=1.5\hbar vq_0$. For small values of $q_0d$,  $-J_{ex}$  increases with increasing separation before reaching the singlet-triplet ground state transition point {$q_0d_c\approx 3$}. For inter-dot distances longer than the critical distance {$d_c\approx 3q_0^{-1}$}, we see the expected exponential decay. Hence in the Klein tunneling regime, the location of the maximum of  $-J_{ex}$ can be roughly predicted by looking at the zero barrier height solution.

\subsection{Effects of well width on two-particle solutions}

\begin{figure}
\includegraphics[trim = 35mm 0mm 35mm 0mm,width=0.36\textwidth]{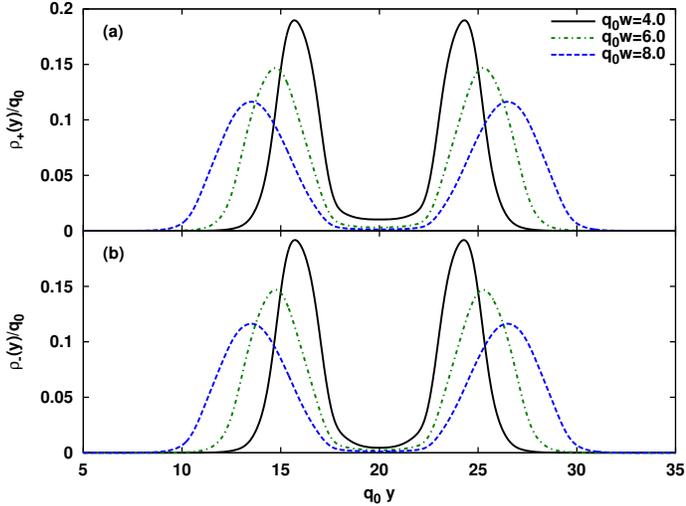}
\caption{Projected charge density $\rho_\pm(y)$ along the ribbon with $q_0d=4$ and $\mu_b=1.9\hbar vq_0$ for (a) singlet state and (b) triplet state for various well widths. $q_0w=4$ (red dotted), $q_0w=6$ (green dash-dot), $q_0w=8$ (blue dashed).   \label{fig6}}
\end{figure}

\begin{figure}
\includegraphics[trim = 30mm 0mm 40mm 0mm,width=0.45\textwidth]{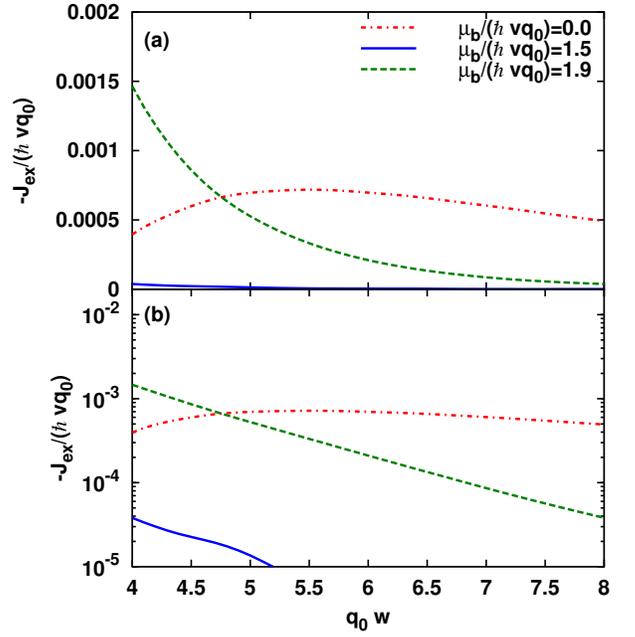}
\caption{Negative exchange coupling ($-J_{ex}$) as a function of well width $w$ for various barrier heights in (a) linear scale and (b) semi-log scale. \label{fig7}}
\end{figure}

Figure~\ref{fig6} shows the charge density along the y-axis for various well widths for (a) the singlet states and (b) the triplet states. The inter-well separation is fixed at $d=4q_0^{-1}$ and the barrier height is $\mu_b=1.9\hbar vq_0$. For small well width with $q_0w=4$ (black solid), the density in the barrier region for the singlet state is much higher than that of the triplet state. For larger well widths, the charge densities spread out from the center, and the difference of the charge densities between the singlet state and the triplet state in the barrier region become less significant. {The absolute value of exchange coupling is hence expected to be very small for large well width.} This is shown in Fig.~\ref{fig7}, where $-J_{ex}$  is plotted as functions of well width. For barrier height larger than the critical value $\mu_b=1.5\hbar vq_0$ (blue solid) and $\mu_b=1.9\hbar vq_0$ (green dashed), the exchange splitting decays exponentially, as shown in the semi-log scale  in Fig.~\ref{fig7}(b). {The exchange splitting of $\mu_b=1.9\hbar vq_0$ is much larger than that of $\mu_b=1.5\hbar vq_0$, since Klein tunneling emerges as the VBM of the barrier  gets close to the conduction band states of the wells.} For zero barrier height (red dash-dot), there is only a single confining well, and the curve is just the long-distance extension of the same curve (red dash-dot) in Fig.~\ref{fig5}(a) discussed in the previous section.

\subsection{Qubit operation}

One figure of merit for qubit operation is the ratio of coherence time to switching time ${T_c}/{\tau_s}$, where $T_c$ is the coherence time and $\tau_s$ is the switching time required for a $\mbox{SWAP}^{\frac{1}{2}}$ gate operation. \cite{LossDiVincenzo1998,Shi2014}Within our model and range of parameters, the maximum exchange splitting for graphene nanoribbon quantum dot qubit is $|J_{ex}|_{graphene}\approx 0.002 \hbar v q_0=66 \mu eV$, which is of the same order as the typical value for GaAs quantum dot qubit $|J_{ex}|_{GaAs}\approx 100 \mu eV$. \cite{Burkard2000,Schliemann2001,Engel2004} Various sources of spin decoherence in graphene quantum dot are investigated, including spin-orbit coupling, \cite{Kane2005,Min2006,Struck2010,Hachiya2014} electron-phonon interaction,  \cite{Mariani2008,Struck2010,Tikhonov2014} and hyperfine interaction. \cite{Fischer2009,Recher2010,Fuchs2012}. The coherence time for graphene is expected to be $T_c \approx 80 \mu s$, three orders of magnitude longer than that of GaAs. \cite{Recher2010,Kloeffel2013} Since the coherence time is much longer and the switching time is similar, we expect the figure of merit of graphene is much better than that of GaAs. In the original proposal, the exchange coupling is estimated to be $J_{ex}\approx 0.1 \sim 1.5 meV$ using single-particle picture and empirical value for Coulomb interaction, \cite{trauzettel2007spin} which is an overestimation comparing to our calculation. Our calculation provides details of the exchange splitting versus various parameters associated with gate operation, and confirms that the magnitude of $J_{ex}$ required for qubit operation is achievable in the presence of electron-electron interaction, even though it is somewhat smaller than the previous estimation.

{
The barrier height of the outside region $\mu_{out}$ should be carefully chosen to avoid the electron leakage caused by Klein tunneling. $\mu_{out}=1.5\hbar vq_0$ in the current model, so an electron state will be confined in the double well if its single-particle energy lies in the band gap of outside region ($0.5\hbar vq_0 < E <2.5\hbar vq_0$). However, if the single-particle energy is too close to the VBM ($0.5\hbar vq_0$) or the bottom of conduction band ($2.5\hbar vq_0$) of the outside region, the electrons could also tunnel out from the double well. In our numerical tests, it is safe to set $\mu_{out}=1.5\hbar vq_0$  and the Fermi energy  $E_F$ is set to be slightly above the conduction band minimum ($1\hbar vq_0$). This region of operation could be identified by measuring the charge stability diagram of a graphene double quantum dot device.}

\section{Conclusion}

We have performed theoretical studies of the electronic structures of GNB quantum dot qubits using the Dirac equation and a double square well potential. The two electron wave functions and exchange splitting are calculated for various potential configurations by using a GVB wave function within the unrestricted Hartree-Fock approximation. As the barrier height approaches $2\hbar v q_0$ (the band gap of the nanoribbon), the magnitude of the exchange coupling is enhanced by the Klein tunneling. This enhancement can make the long distance coupling almost as large as the short distance coupling. {We found that the exchange coupling between two GNB quantum dot qubits can switch sign as the average inter-particle distance varies. This behavior is consistent with previous first-principle studies for two electrons in a finite-length GNB (which corresponds to the zero barrier limit of our GNB double dot system).\cite{Rayne2011} For higher barriers, the magnitude of the exchange coupling decays, but it can be magnified by the Klein tunneling when the valence band maximum of the barrier is close to the conduction band states of the wells. We found that the magnitude for the exchange splitting required for qubit operation is achievable, and the figure of merit of the GNB qubit is expected to be significantly better than that of the GaAs quantum dot qubit.}

\begin{acknowledgments}
We thank Lance Cooper and Jason Chang for valuable discussions. This work was supported in part by Ministry of Science and Technology, Taiwan under Contract No. NSC 101-2112-M-001-024-MY3.
\end{acknowledgments}


\bibliography{gqd}

\end{document}